# Continuous Affect Prediction using Eye Gaze

Jonny O' Dwyer, Ronan Flynn, Niall Murray
Department of Electronics & Informatics
Athlone Institute of Technology,
Athlone, Ireland
j.odwyer@research.ait.ie, rflynn@ait.ie, nmurray@research.ait.ie

*Abstract*—In recent times, there has been significant interest in the machine recognition of human emotions, due to the suite of applications to which this knowledge can be applied. A number of different modalities, such as speech or facial expression, individually and with eye gaze, have been investigated by the affective computing research community to either classify the emotion (e.g. sad, happy, angry) or predict the continuous values of affective dimensions (e.g. valence, arousal, dominance) at each moment in time. Surprisingly after an extensive literature review, eye gaze as a unimodal input to a continuous affect prediction system has not been considered. In this context, this paper evaluates the use of eye gaze as a unimodal input to a continuous affect prediction system. The performance of continuous prediction of arousal and valence using eye gaze is compared with the performance of a speech system using the AVEC 2014 speech feature set. The experimental evaluation when using eye gaze as the single modality in a continuous affect prediction system produced a correlation result for valence prediction that is better than the correlation result obtained with the AVEC 2014 speech feature set. Furthermore, the eye gaze feature set proposed in this paper contains 98% fewer features compared to the number of features in the AVEC 2014 feature set.

## I. Introduction

Affective computing is an interdisciplinary field of computer science, cognitive science, and psychology. It involves tasks of computational analysis, synthesis, recognition, and prediction of human emotion. Emotion classification and continuous affective dimension prediction are two active areas of research within affective computing. Emotion classification is the task of classifying observed emotions into discrete categories such as happy, sad, angry, positive, or negative emotion classes. Continuous affective dimension prediction is the task of predicting a numerical value for an affective dimension such as valence, arousal or dominance for defined periods of time. Studies such as [1] investigated continuous affect prediction using the valence and arousal affective dimensions. Valence refers to the level of pleasure within an emotion, or, how positive or negative the pleasure or displeasure is respectively. Arousal refers to the level of energy or activation associated with the emotion.

High-quality audio-visual databases, such as [2-4], provide baseline feature sets for the speech and video modalities with annotated (labelled in time for arousal and valence) speech and visual recordings. These have been used to build and train models to predict the emotion class or affective dimensions of unknown test speech and video. One of the central issues in affective computing is to define feature sets, extracted from the modality of interest, that best capture the emotion content of that modality. Features sets such as GeMAPS [5], ComParE [6] and AVEC 2014 [3] have been proposed for emotion prediction from speech, all with varying degrees of success.

Eye gaze is defined as the line of sight between one's eyes and an object of fixation [7]. Several studies, as outlined in [8], have shown eye gaze to be critical for both emotion communication and recognition between humans. Interestingly no works investigate eye gaze as a modality to assess emotion on a continuous basis. Eye gaze feature sets, such as that used in [2], were recorded using specialised equipment in a laboratory setting. It was suited to predict emotion after a video sequence had been played and not on a continuous prediction basis. However, the feature set presented in [2], which focused on emotion categorisation, has provided a basis for work on continuous affect prediction presented in this paper.

This paper presents the first unimodal study of eye gaze for the continuous prediction of human emotion, based on the processing of video content using the OpenFace software application [9]. It employs the AVEC 2014 [3] audio-visual database and speech feature set for performance evaluation. This paper proposes a feature set for eye gaze for use in the continuous affect prediction of arousal and valence.

The layout of this paper is as follows: Section II describes related work specific to eye gaze and recent affect recognition experiments. The experimental set-up used for the work presented here is described in Section III. The results are given in Section IV and this is followed in Section V with discussion and comparison with a feature set from the literature. Concluding remarks in Section VI close the paper.

## II. Related Work

This section reviews related work in the areas of eye gaze classification and multimodal affect recognition.

### A. Eye Gaze Classification and Affect Recognition

Unimodal affect recognition using eye gaze was investigated in [10] using an EyeLink 1000 eye tracking device [11]. Eye gaze data was gathered with the eye tracker and processed using neural network models. The highest performing model from this experiment correctly recognized emotion classes as either positive, negative or neutral with an average accuracy of 72.1%, however, only a small population of four males was used.

Automatic eye gaze classification was the focus of [12] for the purpose of determining whether infants were looking or not looking at their parent. The system proposed in [12] included multiple cameras. The data processing element was a support vector machine-based classifier. The findings in [12] suggest that eye gaze direction is important in human-to-human communication. Similar results were presented in [8] and [13].

For human-to-human communication, psychological research suggested a correlation of direct eye gaze with angry and happy emotions [13]. The authors in [13] also claimed that sadness and fear are associated with averted gaze. A geometrical eye and nostril model was used to identify averted gaze and direct gaze on video input. However, in [13] facial illumination of subjects is controlled, which limits the applicability of the results to more natural environments.

Eye gaze behaviour under emotional feedback was studied in [14]. During the experiments, users were asked to watch emotional video sequences and rate the arousal and valence that they perceived from the video. Emotional feedback (correct, incorrect, and random) was provided to the user, in the form of an on-screen emoticon, while they were observing the video. The results showed that random stimuli do not influence a user's emotional state. The emotion recognition system used to assess the eye gaze emotional content used a support vector machine with a radial basis function. For correct and random feedback, the system was reported to have identified the correctly reported arousal 82% of the time compared with 74% for ground-truth. Valence was predicted correctly with only an absolute performance drop of 5% compared to the 75% reported ground truth for this emotional dimension.

### B. Multimodal Affect Recognition

The authors in [2] created an audio-visual affect database (MAHNOB-HCI) and investigated arousal and valence recognition using speech, eye gaze, EEG, and physiological signals in an emotion recognition experiment. The authors divided arousal into classes of medium aroused, calm, and excited for the emotion recognition experiment. Valence was divided into classes of unpleasant, neutral valence, and pleasant. The emotion recognition results from [2] showed that eye gaze performed best during unimodal affect recognition experiments and a combination of eye gaze and EEG proved best overall. It must be stated however, that the speech modality may not have been maximally utilized as the subject was required to watch emotion provoking video only. The eye gaze results for [2] were 63.5% and 68.8% classification accuracy for arousal and valence respectively

A multimodal emotion recognition system was presented in [15], which aimed to classify speech and facial signals into categories of happy, sad, fear, surprise, anger or disgust. For the facial images, the appearance and geometrical features of the eyes and mouth were used along with prosodic and spectral features from the speech for affect classification. The final bimodal system in [15], using speech and facial features, performed better than either single modality for emotion classification.

TABLE I.
AFFECTIVE EYE GAZE FEATURE LIST (31 FEATURES)

| Data | Features |
|---|---|
| Eye gaze distance (2) | eye gaze approach ratio, average eye gaze approach time in milliseconds |
| Eye scan paths (2) | average scan path length, standard deviation of scan path lengths |
| Vertical and horizontal eye gaze coordinates (24) | average, inter quartile range 1-2, inter quartile range 2-3, standard deviation, skewness, power spectral densities at frequencies [0.011, 0.022, 0.033-0.044, 0.055-0.066, 0.077-0.133]Hz, average of standard deviation of coordinates in each fixation zone, standard deviation of standard deviation of coordinates in each fixation zone |
| Eye closure (3) | average eye close frame count, standard deviation of eye close frame count, skewness of eye close frame count |

The authors of [16] created a multimodal affective corpus (emoFBVP) and emotion recognition system using deep belief neural networks. The emotion categories used for classification were the same as [15] except for the addition of a neutral emotion class. The authors used speech, face (including the eye region but not eye gaze), posture, and physiological data from their subjects to predict which class of emotion the subject was experiencing. The specialized hardware system was comprised of a Microsoft Kinect, Zephyr BioHarness, and multiple audio-visual capture devices. The authors reported that emotion was recognised correctly at a rate of 79.2% on the DEAP [17] database using physiological data and 54.8% on the MAHNOB-HCI database [2].

From the review of related work, much has been achieved investigating eye gaze classification, unimodal eye gaze affect recognition, and affect recognition making use of eye gaze as an input within a multimodal system. Of interest was the affective eye gaze feature set presented in [2], which was the only such feature set found in the literature. However, the continuous prediction of emotion using eye gaze was not addressed in the literature. The work presented in this paper focuses on the continuous prediction of emotion using eye gaze. Based on the experimental results obtained, an eye –gaze feature set for continuous arousal and valence prediction is proposed.

## III. EXPERIMENT DESIGN

This section details the framework used for the experiments. Database selection, data point extraction, feature evaluation, and machine prediction methodologies are each discussed.

### A. Database Selection

Video data from the Freeform task of the AVEC 2014 database [3] was selected for input to the eye gaze affect recognition system. The AVEC 2014 database provides arousal, valence, and dominance emotional dimension annotations for video sequences, and ground-truth ratings for these three dimensions. Use of the database allows for the design of continuous affect prediction models for arousal, valence and dominance. A speech feature set is available for the AVEC 2014 tasks and this allows for performance comparison with the proposed eye gaze feature set. The arousal and valence dimensions from the database subset were selected for the experiment prediction tasks.

### B. Raw Data Point Extraction

OpenFace [9] is an open source facial recognition software tool. It can extract raw data points from each frame within a video sequence. In this work, it was used to extract eye gaze data from the AVEC 2014 Freeform dataset. Fig. 1 illustrates the data points used. Additional metadata extracted from the raw data include eyelids opened/closed, eye gaze approach (distance change from user to computer screen), and whether the eyes were scanning or fixed.

### C. Eye Gaze Feature Extraction

Mathematical and statistical features of eye gaze data were computed from overlapping 3 second segments of video frames, moved forward at a rate of 2 seconds for the duration of a video sequence. This method is similar to the short segmentation method used in [3]. The list of features extracted from the data points within each video segment are listed in Table I.

### D. Machine Prediction

Affect dimension prediction was implemented using the SMOreg function for support vector regression in the WEKA data mining software toolkit [18]. The complexity parameter C used to control the model's bias-variance trade-off was set to 0.0325 for valence dimension prediction, and to 0.091 for arousal prediction using eye gaze. C was set to 0.000086 for valence dimension prediction and to 0.000085 for arousal prediction using speech. Based on a series of tests, the C values that resulted in the highest prediction performance were chosen.

The AVEC 2014 dataset contained annotated valence values of 0.0, which means that the annotators had observed video segments that were neither pleasant nor unpleasant. This prompted two experiments where 0.0 ground-truth rated valence values were included and then omitted from model training. The removal of the 0.0 annotated valence ratings resulted in better valence prediction accuracy, therefore this approach was followed for experimental model building.

### E. Feature Evaluation

Eye gaze feature evaluation was performed in WEKA [18]

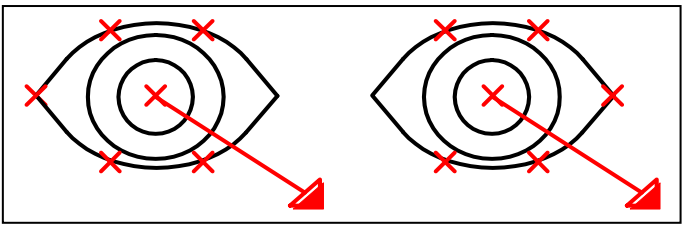

Fig. 1. Eye gaze data points

using two methods. The first was attribute correlation evaluation with a ranker search method and the second was wrapper subset evaluation with a greedy stepwise search. The attribute correlation evaluation method provides information on feature correlation with the annotation target value irrespective of the machine learning scheme used. A correlation ranking is then provided as output ranging from the strongest positive correlation at the top to the strongest negative correlation at the bottom. The wrapper subset evaluation provides ratings for features by using a machine learning scheme, in this case SMOreg. The C values used were the same as for the valence and arousal model training and the attribute selection mode was set to 10-fold cross validation for this task. The feature evaluations carried out in this work provide some preliminary ground for eye gaze feature evaluation and feature engineering tasks in the future.

## IV. RESULTS

### A. Machine Prediction Results for Eye Gaze

The experimental results obtained for the eye gaze affect recognition system are given in Table II. This table details the performance achieved for the system's machine predicted values compared with AVEC 2014 [3] ground truth values, using Pearson's correlation coefficient. The results shown are for performance obtained on the Test partition of the AVEC 2014 Freeform data set. Included in Table II are the correlation results achieved using the AVEC 2014 speech feature set [3] for prediction, which are used for comparison to eye gaze prediction.

### B. Eye Gaze Feature Evaluation

The results of the eye gaze feature evaluation carried out in WEKA can be seen in Tables III and IV. The results in Table III show a Pearson's correlation between a specific feature and the prediction target value. Table III shows that *standard deviation of eye close frame count* achieves the highest correlation score with the target value for both valence and arousal dimensions. Table III also shows that *Y gaze skewness* has one of the strongest negative correlations for both valence and arousal target values. Table IV shows the top ten ranking features after feature set evaluation using the SMOreg learning scheme within the WEKA wrapper subset evaluation [18].

TABLE II.
AFFECT PREDICTION RESULTS FOR EYE GAZE COMPARED WITH THE AVEC 2014 SPEECH FEATURE SET

| Modality | Valence | Arousal |
|---|---|---|
| Eye gaze | 0.315 | 0.329 |
| Speech | 0.308 | 0.512 |

## V. DISCUSSION

Table II shows that the result for continuous valence prediction using eye gaze is better than that of the AVEC 2014 speech feature set, which is chosen as a baseline for comparison. Of significance is that the feature count for the eye gaze affect prediction system presented contains just 31 features compared with the 2,268 features of the AVEC 2014 speech system [3]. A comparison of the figures in Table II for arousal shows that the correlation figure when using eye gaze is less than that obtained for speech using the AVEC 2014 feature set. However, this work is in its initial stages and further improvements for arousal using eye gaze are envisaged using an expanded feature set, which will still be significantly smaller than the 2,268 AVEC 2014 feature set. The decision to use eye gaze for continuous affect prediction is justified by the results obtained for valence, which is the more difficult to predict emotion dimension [14].

The results displayed for the feature evaluation in Table III are interesting as larger negative correlations than positive correlations for the valence dimension can be seen. The results also show a large quantity of negatively correlated features for the valence dimension compared to that of arousal. As can be observed in Table II however, the eye gaze affect recognition system's performance is reasonably similar for both dimensions (Valence = 0.315, Arousal = 0.329). The negatively correlated features within the valence dimension appear to be having a larger effect on the SMOreg algorithm than the positively correlated features. This is supported by the stronger negative correlations for this dimension and by the machine prediction scheme-based feature ranking of Table IV. For example, valence's number one ranked feature from Table III has a correlation of 0.06421 compared with its lowest ranked feature which has a correlation of -0.1445. However, the lowest ranked feature from Table III, *average scan path length*, ranks second place for valence in Table IV where the machine prediction scheme is used to select the feature. These results, specifically the valence features, highlight properties of valence that require further exploration.

TABLE III.
INDEPENDENT EYE GAZE FEATURE CORRELATION TO TARGET VALUE RANKING

| Rank | Features | | | |
|---|---|---|---|---|
| | *Arousal* | *Correlation* | *Valence* | *Correlation* |
| 1 | standard deviation of eye close frame count | 0.22826 | standard deviation of eye close frame count | 0.06421 |
| 2 | X gaze standard deviation | 0.09218 | average eye gaze approach time in milliseconds | 0.0625 |
| 3 | average of standard deviation of X gaze coordinates in each fixation zone | 0.08791 | average eye close frame count | 0.04855 |
| 4 | X gaze inter quartile range 2-3 | 0.0823 | X gaze skewness | 0.04129 |
| 5 | Y gaze inter quartile range 2-3 | 0.08067 | eye gaze approach ratio | 0.04067 |
| 6 | Y gaze standard deviation | 0.07681 | X gaze average | 0.01527 |
| 7 | average scan path length | 0.0727 | X gaze spectral power density 0.011 Hz | 0.00204 |
| 8 | X gaze inter quartile range 1-2 | 0.06756 | X gaze inter quartile range 2-3 | -0.00156 |
| 9 | standard deviation of scan path lengths | 0.06736 | Y gaze average | -0.0028 |
| 10 | average of standard deviation of Y gaze coordinates in each fixation zone | 0.06483 | X gaze spectral power density 0.022 Hz | -0.02119 |
| 11 | X gaze skewness | 0.06447 | X gaze standard deviation | -0.02149 |
| 12 | Y gaze inter quartile range 1-2 | 0.06014 | average of standard deviation of X gaze coordinates in each fixation zone | -0.02289 |
| 13 | standard deviation of standard deviation of X gaze coordinates in each fixation zone | 0.05467 | Y gaze inter quartile range 1-2 | -0.02347 |
| 14 | X gaze spectral power density 0.011 Hz | 0.05449 | Y gaze inter quartile range 2-3 | -0.02597 |
| 15 | X gaze power spectral density 0.055-0.066 Hz | 0.05249 | X gaze spectral power density 0.033-0.044 Hz | -0.02965 |
| 16 | X gaze power spectral density 0.077-0.133 Hz | 0.05001 | standard deviation of standard deviation of X gaze coordinates in each fixation zone | -0.03259 |
| 17 | Y gaze power spectral density 0.077-0.133 Hz | 0.04448 | Y gaze power spectral density 0.022 Hz | -0.03532 |
| 18 | Y gaze power spectral density 0.022 Hz | 0.0406 | X gaze inter quartile range 1-2 | -0.03882 |
| 19 | X gaze power spectral density 0.055-0.066 Hz | 0.03932 | X gaze power spectral density 0.077-0.133 Hz | -0.04068 |
| 20 | X gaze power spectral density 0.022 Hz | 0.03866 | X gaze power spectral density 0.055-0.066 Hz | -0.0459 |
| 21 | eye gaze approach ratio | 0.03562 | Y gaze spectral power density 0.011 Hz | -0.05189 |
| 22 | average eye gaze approach time in milliseconds | 0.03286 | skewness of eye close frame count | -0.05693 |
| 23 | Y gaze spectral power density 0.033-0.044 Hz | 0.03098 | Y gaze spectral power density 0.033-0.044 Hz | -0.05807 |
| 24 | Y gaze power spectral density 0.055-0.066 Hz | 0.0285 | average of standard deviation of Y gaze coordinates in each fixation zone | -0.06451 |
| 25 | Y gaze spectral power density 0.011 Hz | 0.02592 | Y gaze power spectral density 0.055-0.066 Hz | -0.06467 |
| 26 | standard deviation of standard deviation of Y gaze coordinates in each fixation zone | 0.02344 | Y gaze power spectral density 0.077-0.133 Hz | -0.07149 |
| 27 | average eye close frame count | 0.0217 | Y gaze standard deviation | -0.07551 |
| 28 | X gaze average | -0.00927 | standard deviation of standard deviation of Y gaze coordinates in each fixation zone | -0.08062 |
| 29 | skewness of eye close frame count | -0.04429 | Y gaze skewness | -0.09209 |
| 30 | Y gaze skewness | -0.06792 | standard deviation of scan path lengths | -0.09662 |
| 31 | Y gaze average | -0.14595 | average scan path length | -0.10497 |

TABLE IV.
SMOREG MACHINE PREDICTION SCHEME BASED FEATURE RANKING

| Rank | Arousal Features | Valence Features |
|---|---|---|
| 1 | standard deviation of eye close frame count | Y gaze skewness |
| 2 | Y gaze average | average scan path length |
| 3 | Y gaze skewness | standard deviation of standard deviation of Y gaze coordinates in each fixation zone |
| 4 | X gaze skewness | X gaze skewness |
| 5 | Y gaze inter quartile range 2-3 | average eye gaze approach time in milliseconds |
| 6 | average eye gaze approach time in milliseconds | standard deviation of eye close frame count |
| 7 | skewness of eye close frame count | eye gaze approach ratio |
| 8 | standard deviation of scan path lengths | skewness of eye close frame count |
| 9 | X gaze standard deviation | X gaze spectral power density 0.055-0.066 Hz |
| 10 | standard deviation of standard deviation of X gaze coordinates in each fixation zone | Y gaze spectral power density 0.055-0.066 Hz |

## VI. CONCLUSION

This paper presented a study of eye gaze as a unimodal input to a continuous affect prediction system. An affective eye gaze feature set was presented for the continuous prediction of valence and arousal emotion dimensions. The performance of eye gaze as an input to an affect prediction system was compared to using speech as an input. Better results were achieved for valence prediction when using eye gaze, with a much smaller feature set. The proposed eye gaze feature set is 98.64% smaller than the speech feature set used for comparison (31 eye gaze features compared with 2,268 speech features). Eye gaze did not perform as well as speech for arousal prediction, a correlation value of 0.329 for eye gaze compared with 0.512 for speech. However, the experimental results from this study will form the basis of further research for continuous affect prediction using eye gaze features. Future work will include further eye gaze feature engineering, the refinement of the feature set presented in this paper, and the evaluation of this study's features for use with other machine learning schemes. It is also intended that the findings in this paper be extended to trials on other audio-visual affect databases. The inclusion of eye gaze in multimodal affect prediction systems will also be investigated.

ACKNOWLEDGEMENT

This work was supported by the Irish Research Council under grant number GOIP/2016/1572.

REFERENCES

[1] M. A. Nicolaou, H. Gunes, and M. Pantic, "Continuous Prediction of Spontaneous Affect from Multiple Cues and Modalities in Valence-Arousal Space," *IEEE Transactions on Affective Computing*, vol. 2, no. 2, pp. 92–105, Apr. 2011.

[2] M. Soleymani, J. Lichtenauer, T. Pun, and M. Pantic, "A Multimodal Database for Affect Recognition and Implicit Tagging," *IEEE Transactions on Affective Computing*, vol. 3, no. 1, pp. 42–55, Jan. 2012.

[3] M. Valstar *et al.*, "AVEC 2014: 3D Dimensional Affect and Depression Recognition Challenge," in *Proceedings of the 4th International Workshop on Audio/Visual Emotion Challenge*, New York, NY, USA, 2014, pp. 3–10.

[4] G. McKeown, M. Valstar, R. Cowie, M. Pantic, and M. Schroder, "The SEMAINE Database: Annotated Multimodal Records of Emotionally Colored Conversations between a Person and a Limited Agent," *IEEE Transactions on Affective Computing*, vol. 3, no. 1, pp. 5–17, Jan. 2012.

[5] F. Eyben *et al.*, "The Geneva Minimalistic Acoustic Parameter Set (GeMAPS) for Voice Research and Affective Computing," *IEEE Transactions on Affective Computing*, vol. 7, no. 2, pp. 190–202, Apr. 2016.

[6] Björn Schuller, Stefan Steidl, Anton Batliner, Julia Hirschberg, Judee K. Burgoon, Alice Baird, Aaron Elkins, Yue Zhang, Eduardo Coutinho, Keelan Evanini: "The INTERSPEECH 2016 Computational Paralinguistics Challenge: Deception, Sincerity & Native Language", *Proceedings INTERSPEECH 2016, ISCA, San Francisco, USA, 2016*.

[7] O. Lappi, "Eye movements in the wild: Oculomotor control, gaze behavior & frames of reference," *Neuroscience & Biobehavioral Reviews*, vol. 69, pp. 49–68, Oct. 2016.

[8] R. J. Itier and M. Batty, "Neural bases of eye and gaze processing: The core of social cognition," *Neuroscience & Biobehavioral Reviews*, vol. 33, no. 6, pp. 843–863, Jun. 2009.

[9] T. Baltrušaitis, P. Robinson, and L. P. Morency, "OpenFace: An open source facial behavior analysis toolkit," in *2016 IEEE Winter Conference on Applications of Computer Vision (WACV)*, 2016, pp. 1–10.

[10] C. Aracena, S. Basterrech, V. Snáel, and J. Velásquez, "Neural Networks for Emotion Recognition Based on Eye Tracking Data," in *2015 IEEE International Conference on Systems, Man, and Cybernetics*, 2015, pp. 2632–2637.

[11] SR Research. Eye Link. Available at: http://www.sr-research.com/EL_II.html, date of access: 16/02/2017

[12] S. Cadavid, M. H. Mahoor, D. S. Messinger, and J. F. Cohn, "Automated classification of gaze direction using spectral regression and support vector machine," in *2009 3rd International Conference on Affective Computing and Intelligent Interaction and Workshops*, 2009, pp. 1–6.

[13] Y. Zhao, X. Wang, and E. M. Petriu, "Facial expression analysis using eye gaze information," in *2011 IEEE International Conference on Computational Intelligence for Measurement Systems and Applications (CIMSA) Proceedings*, 2011, pp. 1–4.

[14] F. Ringeval, A. Sonderegger, B. Noris, A. Billard, J. Sauer, and D. Lalanne, "On the Influence of Emotional Feedback on Emotion Awareness and Gaze Behavior," in *2013 Humaine Association*

Conference on Affective Computing and Intelligent Interaction, 2013, pp. 448–453.
[15] S. Thushara and S. Veni, "A multimodal emotion recognition system from video," in *2016 International Conference on Circuit, Power and Computing Technologies (ICCPCT)*, 2016, pp. 1–5.
[16] H. Ranganathan, S. Chakraborty, and S. Panchanathan, "Multimodal emotion recognition using deep learning architectures," in *2016 IEEE Winter Conference on Applications of Computer Vision (WACV)*, 2016, pp. 1–9.
[17] S. Koelstra, C. Mühl, M. Soleymani, A. Yazdani, J.-S. Lee, T. Ebrahimi, T. Pun, A. Nijholt, and I. Patras, "DEAP: A Database for Emotion Analysis Using Physiological Signals," *IEEE Trans. Affective Computing*, vol. 3, no. 1, pp. 18-31, Jan.-Mar. 2012.
[18] M. Hall, E. Frank, G. Holmes, B. Pfahringer, P. Reutemann, and I. H. Witten, "The WEKA Data Mining Software: An Update.," *SIGKDD Explorations*, vol. 11, no. 1, pp. 10–18, 2009.